\documentclass[12pt]{article}
\usepackage{epsfig}

\def\a{\alpha}

\def\b{\beta}

\def\th{\theta}
\def\Th{\Theta}
\def\S{\Sigma}
\def\s{\sigma}

\def\F{\Phi}

\def\D{\Delta}
\def\e{\epsilon}

\def\G{\Gamma}
\def\g{\gamma}
\def\o{\omega}
\def\O{\Omega}
\def\p{\pi}
\def\d{\delta}
\def\m{\mu}
\def\n{\nu}

\def\L{\Lambda}

\def\det{\textrm{det}}
\def\i{\iota}

\def\cd{{\cal D}}

\def\cj{{\cal J}}

\def\ha{\frac12}
\def\pr{\partial}
\def\to{\rightarrow}

\newcommand{\be}{\begin{equation}}
\newcommand{\ee}{\end{equation}}
\newcommand{\bea}{\begin{eqnarray}}
\newcommand{\eea}{\end{eqnarray}}

\begin{document}

\begin{titlepage}

\bigskip
\bigskip
\bigskip
\bigskip

\begin{center}

{\bf{\Large Flat Spacetime Vacuum in Loop Quantum Gravity}}

\end{center}
\bigskip
\begin{center}
 A. Mikovi\'c \footnote{E-mail address: amikovic@ulusofona.pt}
\end{center}
\begin{center}
Departamento de Matem\'atica e Ci\^{e}ncias de Computac\~ao,
Universidade Lus\'ofona, Av. do Campo Grande, 376, 1749-024,
Lisboa, Portugal
\end{center}

\normalsize

\bigskip
\bigskip
\begin{center}
                        {\bf Abstract}
\end{center}
We construct a state in the loop quantum gravity theory with zero
cosmological constant, which should correspond to the flat
spacetime vacuum solution. This is done by defining the loop
transform coefficients of a flat connection wavefunction in the
holomorphic representation which satisfies all the constraints of
quantum General Relativity and it is peaked around the flat space
triads. The loop transform coefficients are defined as spin foam
state sum invariants of the spin networks embedded in the spatial
manifold for the $SU(2)$ quantum group. We also obtain an
expression for the vacuum wavefunction in the triad represntation,
by defining the corresponding spin networks functional integrals
as $SU(2)$ quantum group state sums.
\end{titlepage}
\newpage

\section{Introduction}

In \cite{mfc} it was demonstrated that the invariants of the
embedded spin networks for the $SU(2)$ BF theory can be
interpreted as the loop transform coefficients of a flat
connection wavefunction which satisfies all the constraints of
quantum General Relativity (GR) in the Ashtekar formulation. Since
the flat spacetime can be represented as a flat connection
solution with the flat metric constraint, this opened up a
possibility of constructing a flat metric vacuum state in the
framework of loop quantum gravity \cite{RSlr,RSsn,lgr}.

The approach of constructing a vacuum state by using the spin
network invariants was initiated by Smolin for the case of
non-zero cosmological constant, see \cite{Smo} for a review and
references. In this case, a solution of the quantum constraints
has been already known, which is the Kodama state \cite{Ko}. The
corresponding loop transform coefficients are given by the
$SU(2,{\bf C})$ Chern-Simons (CS) theory invariants of the
embedded spin networks, and it was conjectured that these
invariants are given by an analytic continuation of the
$SU_q(2)$ quantum group spin networks evaluations
at a root of unity \cite{Smo}.
The quantum group evaluations of the spin networks are given
by the Kauffman brackets \cite{K,KL,qgb}. Consequently, one has to
use the quantum spin networks, i.e. the spin networks where the
$SU(2)$ irreducible representations (irreps) are restricted by
$j\le k/2$, where $k\in \bf N$ and $k=6\p/\L l_P^4$, where $\L$ is
the cosmological constant and $l_P$ is the Planck length.
In the $\L=0$
case, the CS theory is replaced by the BF theory, and in order to
define the corresponding spin network invariants one must use the
same category of the $SU_q(2)$ irreps \cite{mfc}, but then
$k$ is an arbitrary natural number.

The state constructed in \cite{mfc} corresponds to a flat
connection wavefunction which is not peaked around any particular
value of the triads, and hence cannot be a good description of the
flat metric vacuum. The same problem also appears in the case of
non-zero cosmological constant, because the Kodama state is not
peaked around any particular value of the
triads\footnote{Classically, self/anti-self dual connection
solution describes de-Sitter/anti-de-Sitter solution only if the
spatial metric is flat.}.

In this paper we extend the construction of \cite{mfc} to the case
when the flat connection wavefunction is peaked around the flat
metric values of the triads, so that we obtain a state in the spin
network basis which could be considered as a flat space vacuum.
In section 2 we give a review and some basic formulas of the
holomorphic representation. These formulas are then used in
section 3 to construct a flat connection wavefunction which is
peaked around the flat space triads. In section 4 the
corresponding spin network basis coefficients are defined as
the $SU_q(2)$ spin foam state sum invariants. Since the vacuum
wavefunction satisfies the quantum constraints in the holomorphic
representation, in section 5 we construct the wavefunction in the
triad representation by defining the corresponding functional
integrals as $SU_q(2)$ state sums. In section 6 we present our
conclussions, and in the Appendix we present the calculation of
the expansion coefficients for the gauge invariant plane waves.

\section{Holomorphic representation}

The Ashtekar formulation of GR \cite{A} is based on the complex
canonical variables \be A_i^a = -i\o_i^a (e) +  \tilde p_i^{a}
\quad,\quad E^i_a = \tilde e^i_a \quad,\label{ashv}\ee where
$\tilde e^i_a =\sqrt{h} e^i_a$, $e$'s are the inverse triads, $h$
is the determinant of the metric on the three-manifold $\S$,
$\o(e)$ is the torsion free spin connection and $\tilde p$ is the
canonically conjugate variable to $\tilde e$ \footnote{Here we use
the conventions such that the spin connection is the imaginary
part of the complex connection $A$. Hence the curvature two-form
is given by $F^a = dA^a + i\e_{abc}A^b\wedge A^c $ where the
$SU(2)$ Lie algebra generators $T_a$ satisfy
$[T_a,T_b]=i\e_{abc}T_c$. An $SU(2)$ group element is given by
$g=e^{i\Th^a T_a}$, where $\Th^a$ are real parameters.}. The
constraints of GR become polynomial in terms of the new variables,
and the Hamiltonian constraint is given by \be C_H =
\e_{abc}\left( E^{ia} E^{jb} F^{c}_{ij} + \L\e_{ijk} E^{ia} E^{jb}
E^{kc} \right) \quad,\label{hamc} \ee where $\L$ is the
cosmological constant.

The polynomiality of the constraints simplifies the quantization
of the theory; however, one has to find a representation of the
complex canonical variables which is consistent with the reality
conditions, i.e. \be A^\dag - A = 2i\o (E)\quad,\quad E^\dag = E
\quad.\ee In the case of finitely many canonical variables, this
is resolved via the holomorphic representation \cite{Ko2}. Given a
Heisenberg algebra $[p,q]=i$, one can introduce creation and
annihilation operators \be a = p + i\O^{\prime}(q) \quad,\quad
a^\dag = p -i \O^{\prime}(q) \quad,\ee where $\O$ is a given
function of $q$ and $\O^{\prime}={d\O\over dq}$. Since the
operators $a$ and $a^\dag$ are not hermitian, their eigenvalues
$\a$ and $\bar\a$ will be complex numbers, where $\bar\a$ is the
complex conjugate of $\a$. Let us for the sake of convinience
denote the eigenvector of $a^\dag$ as $|\a\rangle$, so that \be
a^\dag |\a\rangle = \bar\a |\a\rangle \quad.\ee These eigenstates
can be expressed in the $q$ representation as \be\langle q | \a
\rangle = e^{-i\bar\a q + \O (q)} \quad,\ee so that for an
arbitrary state $|\Psi\rangle$ we can define $\Psi(\a)$ as \be
\Psi (\a) = \langle \a|\Psi \rangle = \int
dq\,\langle\a|q\rangle\langle q|\Psi\rangle = \int dq \, e^{i\a q
+ \O (q)}\F (q) \quad.\ee

The complex variable function $\Psi(\a)$ will be holomorphic, and
one can define the holomorphic representation as
\be a|\Psi\rangle \to \a \Psi (\a) \quad,\quad
q|\Psi\rangle \to i\Psi^{\prime}(\a)\quad.\ee
Given a holomorphic wavefunction $\Psi(\a)$, one can go back to the
$q$ representation via the contour integral
\be \F (q) = e^{-\O (q)}\int_{C=\bf R} d\a\, e^{-i\a q} \Psi(\a)
\quad,\ee
so that in order to obtain the usual wavefunction we need to know
$\Psi(\a)$ for the real values of $\a$.

In the case of GR one can define the holomorphic representation as
\cite{Ko2} \be \hat A |\Psi\rangle \to A \Psi[A] \quad,\quad \hat
E |\Psi\rangle \to i{\d \Psi\over \d A} \quad,\ee where \be \Psi
[A] = \langle A|\Psi \rangle = \int \cd E \,\langle
A|E\rangle\langle E|\Psi\rangle = \int \cd E \, e^{i\int_{\S}d^3 x
\,A E + \O [E]}\F [E] \quad, \label{psi}\ee and ${\d\O\over\d E} =
\o (E)$. Given a $\Psi[A]$ functional, we can obtain the $\F [E]$
functional as \be \F [E] = e^{-\O [E]}\int_{A\in \bf R} \cd A \,
e^{-i\int_\S d^3 x A E} \Psi[A] \quad.\label{fi}\ee Although the
expresions in the GR case are a straightforward generalization of
the finite-dimensional ones, the key difference is that all the
integrals that are used in the finite-dimensional case become
functional integrals, and these have to be defined. For our
purposes it will suffice to define the functional integral
(\ref{fi}).

\section{Flat-connection wavefunctions}

In \cite{mfc} it was pointed out that the wavefunctions $\d
(F)\Psi_0 (A)$ are solutions of the $\L = 0$ GR constraints in the
holomorphic representation. Since at the classical level the $F=0$
solutions include the flat metric solution, i.e. when $A=i\o(E)$
then $F(A)=iR_3$, where $R_3$ is the scalar curvature of $\S$. One
can then use the $\d (F)\Psi_0 (A)$ wavefunctions to construct a
flat metric vacuum state \cite{mfc}.

Let $|0\rangle$ be a state
corresponding to $\d (F)\Psi_0 (A)$, then one can represent this
state in the basis of diffeo invariant spin network states
$|\g\rangle$ as \cite{RSsn}
\be |0\rangle = \sum_\g \langle\g |0\rangle |\g\rangle \quad,\ee
where
\be \langle\g |0\rangle = \int \cd A\, W_\g [A]\d (F)\Psi_0 [A] =I_\g
\quad.\label{snf}\ee
The complex number $I_\g$ is a topological invariant for a spin
network $\g$ embedded into the three-manifold $\S$, and the formal
functional integral expression (\ref{snf}) can be
used to define this invariant. This can be done by replacing the $\S$
by a simplical complex corresponding to a triangulation of $\S$ and
then by using the dual one-complex to define the integral over the
connections
as an integral over the group elements associated to the dual edge
holonomies. This finite-dimensional integral can be then regulated
via the quantum $SU(2)$ group at a root of unity, which can be
represented as a spin foam state sum. This was done in
\cite{mfc} for the case of $\Psi_0 =1$.

Note that in the expression (\ref{snf}) the integration is over
the complex $A$, so that one has to use the $SU(2,\bf{C})$ group
\footnote{The Lie algebra $su(2,\bf{C})$ is isomorphic to the Lie
algebra $sl(2,\bf{C})$, when $sl(2,\bf{C})$ is considered as a
complex vector space. If we consider $sl(2,\bf{C})$ as a real
vector space, then it is isomorphic to $su(2)\oplus su(2)$.}.
Since the relevant category of irreps is always over $\bf C$, and
we do not require unitarity of the quantum group irreps, then the
categories of finite-dimensional irreps  of $SU_q (2,\bf{R})$ and
$SU_q (2,\bf{C})$ are equivalent.

It was also argued in \cite{mfc} that the state $\Psi(A)=\d (F)$
can not correspond to a flat metric vacuum because it does not
favor any particular value of the operator $\hat E$. It was then
suggested to take a nontrivial $\Psi_0 (A)$, which would
correspond to a state which is sharply peaked around the flat
metric value $E_0$ of $\hat E$. Observe that if $\Psi_0 (A)$ is an
eigenstate of $\hat E$, then $\Psi (A) = \d (F) \Psi_0 (A)$ is
not, but we can still consider the state $\Psi(A)$ as a state
which is sharply peaked around $E_0$. The reason is that the
formula (\ref{snf}) can be rewritten as \be \langle\g |0\rangle =
\int \cd A^*\, W_\g [A^*]\Psi_0 [A^*] \quad,\label{rsf}\ee i.e. as
a functional integral over the flat connections $A^*$, see also
\cite{ba}. Hence one can reinterpret $|0\rangle$ as a state from a
vector space corresponding to a quantization of the space of flat
connections. Let $E^*$ be the canonical conjugate of $Re\,A^*$,
then the holomorphic representation is defined by the same
formulas as in the case of a non-flat connection $A$. In
particular we have \bea \Psi_0 [A^*] &=& \langle A^* |\Psi_0
\rangle
= \int \cd E^* \,\langle A^* |E^* \rangle\langle E^* |\Psi_0\rangle \nonumber\\
&=& \int \cd E^* \, e^{i\int_{\S}d^3 x \,A^* E^* + \O [E^* ]}\F [E^* ] \quad,
\label{pfc}\eea
so that if we choose for $\F$ an eigenfunctional of $\hat E^*$ for the
flat-metric eigenvalue $E_0$, which is given formally by $\d (E^* - E_0)$,
we then obtain
\be \Psi_0 [A^*] = e^{i\int_{\S}d^3 x \,A^* E_0 - \O [E_0 ]} =
e^{i\int_{\S}d^3 x \,A^* E_0 } \quad,
\label{psz}\ee
since $\O (E_0) =0$.

Therefore this analysis suggests to take for the flat metric vaccum a state
$|0\rangle$ whose coefficients in the spin network basis are given by the
formal expression
\be \langle\g |0\rangle =
\int \cd A\, W_\g [A]\d (F)e^{i\int_{\S}d^3 x \,A E_0 }
\quad.\label{vin}\ee
In the next section we will define this functional integral as a spin-foam
state sum for the quantum $SU(2)$ group at a root of unity.

\section{Flat metric spin network invariants}

Let $T(\S)$ be a simplical complex corresponding to a
triangulation of $\S$. We consider only the triangulations where
the dual one-complex of $T(\S)$ is a four-valent graph $\G$, see
Fig. 1. We can define (\ref{vin}) by taking \be \cd A = \prod_l
dg_l \quad,\label{im}\ee where $l$ labels the edges of $\G$, $g_l$
is an $SU(2)$ group element corresponding to the edge holonomy and
$dg_l$ is the usual group measure (Haar measure). Then $\d(F)$ can
be defined as \be \prod_f \d (g_f) \quad,\ee where $f$ labels the
faces of $\G$ and $g_f = \prod_{l\in \pr f} g_l$. The group delta
function is defined as \be \d(g) = \sum_{j} (2j +1)\, Tr
D^{(j)}(g) \quad,\ee where $j$ is the spin and $D^{(j)}$ is the
corresponding representation matrix.

We will embedd the spin network $\g$ into the graph $\G$ by
identifying a subset of vertices $V^\prime$ of $\G$ with the
vertices of $\g$ and then we will assign to the edges of $\G$
which connect the verticies from $V^\prime$ the irreps of the spin
network $\g$ (one can have more than one irrep of $\g$ associated
to the same edge of $\G$). Then \be W_\g [A] = \big\langle
\prod_{l\in L^{\prime}}\prod_{j_l \in J_\g } D^{(j_l)}(g_l)
\prod_{v\in V^\prime}C^{(\i_v )}\big\rangle \quad,\label{swl}\ee
where $L^\prime$ is a set of edges of $\G$ labelled by the irreps
of $\g$, $J_\g$ is the set of irreps of $\g$, $C^{(\i)}$ are the
intertwiner tensors for the intertwiners of $\g$ and
$\langle\,\rangle$ denotes the $SU(2)$ trace \cite{PfO}.

As far as $\Psi_0 = e^{i\int_{\S}d^3 x \,\langle A  E_0 \rangle }$
is concerned, it can be discretized as \be \Psi_0 = \exp
i\sum_{l\in L} \langle A_l  E^0_{\D}\rangle = \prod_{l\in
L}e^{i\langle A_l E^0_\D\rangle} \quad, \label{swf}\ee where \be
E^a_\D = \int_\D E^{ai} \e_{ijk}dx^i \wedge dx^j \quad,\ee and
$\D$ is a triangle dual to the edge $l$. Due to the Peter-Weyl
theorem, one has \be e^{iA^a E_a} = \sum_j (c_j (E))^\a_\b
D^{(j)\b}_\a (e^{iA^a T_a})\quad,\label{pve}\ee where \be (c_j
(E))^\a_\b =(2j+1) \int_{SU(2)} dg \bar{D}^{(j)\b}_\a (g)f(g)
\quad,\quad f(g)=f(e^{iA^a T_a})=e^{iA^a E_a}\quad.\ee We then get
\be \Psi_0 = \sum_{j_1,...,j_L} c^{\a_1}_{\b_1} \cdots
c^{\a_L}_{\b_L} D^{(j_1)\b_1}_{\a_1}(g_1) \cdots
D^{(j_L)\b_L}_{\a_L}(g_L) \quad.\ee

Note that in order to have a good definition of $I_\g$, the
function $\Psi_0$ should be gauge invariant. This translates into
the invariance under $g_l \to h^{-1}_v g_l h_{v^\prime}$, where $v$
and $v^\prime$ are the ends of $l$ and $h$'s are arbitrary group
elements. This invariance requires that one finds an approximation
\be c^{\a_1}_{\b_1} \cdots
c^{\a_L}_{\b_L} \approx \sum_{\i}f(E,\i)C^{\a_1 ... \a_L (\i)}_{\b_1 ...
\b_L} \quad,\label{cappr}\ee where $C^{(\i)}$ are the intertwiner tensors
\footnote{One cannot have an equality because the coefficients
$c_\a^\b (E)$ are not the group tensors.}. The
simplest way to achieve this is to replace the plane-vawe
expansion (\ref{pve}) by a gauge-invariant expression \be
(e^{iAE})_{inv} = \sum_j C_j(E) Tr D^{(j)}(e^{iAT})\quad,
\label{gipv}\ee where $(2j+1)C_j (E) = Tr\,(c_j (E))$. Then by taking
the trace of both sides of the approximation (\ref{cappr}), and assuming
that all the $f$'s are equal, one obtains
\be f(E,\i) = {C_{j_1}(E_1) \cdots C_{j_L}(E_L)
\over (2j_1 +1)\cdots (2j_L + 1)}  \quad.\ee

It is not
difficult to calculate the coefficients $C_j$, see the Appendix.
An interesting feature is that they are non-zero only for $j\le
\e=|E|/l^2_{P}$. The $C_j$'s are given
by the expression \be C_j (E) = \left[ I_0 (j)-I_0
(j+1) + I_2 (j)-I_2 (j+1) \right] \quad, \label{cj}\ee where \be
I_0 (j)= {\th(\e -j)\over\sqrt{\e^2 - j^2 }} \quad,\quad I_2 (j)=
{1-2j^2 /\e^2 \over \sqrt{\e^2 - j^2} }\th(\e -j)
 \quad,\ee
where $\th$ is the step function ($\th(x)=0$ for $x\le 0$ and
$\th(x)=1$ for $x>0$).

In the case when $(E_0)_i^a = \d_i^a$ (flat space), it is natural to take $\e =1$,
so that the expansion (\ref{gipv}) terminates at $j=1$. The $C_1$ coefficient is
then divergent, and one would have to decide how to regularize it. The corresponding
$I_\g$ will be given by the integral
\be I_\g =\sum_{j^\prime_l ,\i^\prime }  \int \prod_l dg_l \tilde C_{j^\prime_l}(E_0)
\prod_f \d (g_f)
W_\g (g_{l^\prime},j,\i ) \langle C^{(\i^\prime )}D^{(j^\prime_1)}(g_1)\cdots
D^{(j^\prime_L)}(g_L)
\rangle\quad,\ee
where $(2j+1)\tilde C_j = C_j$. This integral
can be represented by a state sum
\be I_\g = \sum_{j_f , \i_l ,j^\prime_l ,\i^\prime_v}\prod_f (2j_f +1)\prod_l
\tilde C_{j^\prime_l}(E_0)
 \prod_v A_v (j_f, \i_l ; j^\prime_l ,\i^\prime_v ; j, \i ) \quad.\label{css}\ee

The vertex amplitudes $A$ are evaluations of the vertex spin
networks associated to the spin foam $\{\G,j_f,\i_l \}$ carrying
the spin networks $\{\G,j^\prime_l ,\i^\prime_v \}$ and
$\{\g,j_{l^\prime} ,\i_{v^\prime} \}$. The vertex spin networks
can be obtained in the following way: draw the circuit diagram
$\G_c$ for $\G$ (each circuit is labelled by a $j_f$) and then
draw the $\G(j^\prime,\i^\prime)$ and the $\g(j,\i)$ spin networks
on the $\G_c$ diagram. This gives the circuit diagram $\G_c
(\G,\g)$, see Fig. 2. Then shrink each edge of $\G_c (\G,\g)$ to a
point, and label these points with the intertwiners $\i_l$. These
rules follow from the graphical representation of the group
integrations \cite{fksf}, see Fig. 3. In this way one obtains the
vertex spin networks which are the tetrahedral spin networks with
the additional four-vertices $j^\prime,\i^\prime$ and additional
lines and vertices corresponding to the spin network $\g$, see
Fig. 4.

The state sum for $I_\g$ is regularized by passing to the category
of finite dimensional irreps of $SU_q (2)$ where $q$ is a root of
unity. Let $q=e^{2\pi i/(k+2)}$, $k\in \bf N$, then the $SU(2)$
irreps satisfy $j\le k/2$ and \be I_\g = \sum_{j_f ,\i_l
,j^\prime_l ,\i^\prime_v} \prod_f \dim_q (j_f)\prod_l
C_{j^\prime_l}(E_0)\prod_v
 A_v^{(q)}(j_f, \i_l ; j^\prime_l ,\i^\prime_v ; j, \i ) \quad,\label{ssi}\ee
where $\dim_q$ is the quantum dimension and $A_v^{(q)}$ is the quantum
group evaluation of the vertex spin network. These spin networks evaluations can be
calculated by using the formulas from \cite{qgb}.

\section{The triad representation}

Given the invariants $I_\g$ for the flat-space vacuum, one can
construct the corresponding wavefunction as \be \Psi [A] = \sum_\g
I_\g \, \bar W_\g [A] \quad.\ee Since $A$ is complex, we want to
obtain the usual wavefunction, i.e. the function of the real
argument. We then use the formula (\ref{fi}) to go to the triad
representation, so that \be \F [E] = \sum_\g I_\g \bar\F_\g [E]
\quad,\label{fe}\ee where \be \F_\g [E] = e^{-\O [E]}\int_{A\in
\bf R} \cd A \, e^{-i\int_\S d^3 x A E} W_\g [A] \quad.\label{wle}
\ee

This functional integral can be also defined as a state sum via
the simplical decomposition of $\S$. As in the previous section we
use the formulas (\ref{im}), (\ref{swl}) and (\ref{swf}) so that
\be \F_\g [E] = \int\prod_l dg_l \prod_{l\in L}e^{-iA_l E_\D}
\big\langle \prod_{l\in L^{\prime}}\prod_{j_l \in J_\g }
D^{(j_l)}(g_l) \prod_{v\in V^\prime}C^{(\i_v )}\big\rangle\quad.
\ee By using the gauge invariant approximation (\ref{cappr}) we
obtain the state sum \be \F_\g [E]=
\sum_{j^\prime,\i^\prime,\tilde\i}\,C_{j^\prime_1} (E_1)\cdots
C_{j^\prime_L} (E_L)  \prod_{v \in V^\prime} \cj^{(v)}_{\G,\g}
(j^\prime ,\i^\prime ;j,\i;\tilde\i) \quad,\label{ess}\ee where
$\cj^{(v)}_{\G,\g}(j^\prime ,\i^\prime ;j,\i;\tilde\i)$ are
evaluations of the vertex spin networks obtained by a composition
of the spin network $\{\G, j^\prime,\i^\prime \}$ and the spin
network $\{\g,j,\i \}$.

The vertex spin networks can be obtained by using the following
rules: draw the graph of the spin network $\g$ close to the graph
of the spin network $\G$, such that the resulting graph
corresponds to the embedding of the spin network $\g$ into the
one-complex $\G$, see Fig. 5. Then use the graphical rules for the
group integration, see Fig. 3, so that each edge of the $\G
\bigcup \g$ graph is cut into two vertices. These new vertices
will carry the intertwiners $\tilde\i$, if they are four-valent or
higher. If they are two-valent vertices, there will be a factor of
the inverse dimension of the corresponding $\g$ irrep. If the two
new vertices are one-valent (this happens in the case when an edge
of $\G\bigcup\g$ carries only an irrep of the spin network $\G$)
such vertices are not included, because they will carry the
trivial identity irrep. In this way one obtains non-trivial spin
networks only for the vertices of $\G$ which are close to the
vertices of $\g$, see Fig. 6.

The state sum (\ref{ess}) can be then regularized by using the same
category of the finite-dimensional irreps of the $SU_q(2)$ quantum group
as in the case of the invariant $I_\g$, i.e. $q=e^{2\pi i/(k+2)}$, $k\in
\bf N$ and $j\le k/2$.

\section{Conclussions}

We expect that the state sum (\ref{ssi}) for $I_\g$ should be a
topological invariant because it was based on the diffeomorphism
invariant formal expressions (\ref{vin}) and (\ref{css}). This
means that $I_\g$ should be independent of the choice of a
triangulation of $\S$, which means independence of the dual
one-complex $\G$. Still, one should check the triangulation
invariance because it may require different then the usual
normalization factors in the spin network amplitudes. This will
require checking the invariance under the Pachner moves, see
\cite{sfst} for the case when $\S$ is a two-dimensional manifold
and $\Psi_0 =1$.

Note that the state sum $I_\g$ corresponds to a new type of a spin
foam, due to the presence of a nontrivial function $\Psi_0$.
Beside labelling the faces of the two-complex $\G$ with arbitrary
group irreps, we also have to label the edges of $\G$ with
arbitrary group irreps.

The expression for the wavefunction in the triad representation
(\ref{fe}) will have the form \be\Phi [E] = \sum_\g f_{\g,\G}
(E_1,...,E_L)\quad.\label{fs}\ee Since the number of vertices of a
spin network $\g$ satisfies $V_\g \le V_\G$, this means that the
expression for $\Phi[E]$ will involve a sum over different $\G$'s,
or equivalently a sum over different triangulations of $\S$. Note
that summing over triangulations is a way of obtaining
diffeomorphism invariant expressions when the individual terms are
not diffeomorphism invariant. This raises a possibility to
approximate $\Phi$ by taking a sufficiently big graph $\G$, i.e. a
sufficiently fine triangulation of $\S$, which would then truncate
the infinite sum (\ref{fs}) to a finite sum over the $\g$'s such
that $V_\g \le V_\G$.

Given the functional $\Phi [E]$, the central question is whether
it has a good semi-classical limit. This amounts to showing that
an apropriately defined semi-classical spacetime metric
$g_{\m\n}(t,x)$ ($t\in{\bf{R}}\,,\,x\in\S$) satisfies quantum
corrected Einstein equations \be R_{\m\n} + l_P c_1 (\nabla
R)_{\m\n} + l_P^2 [c_2 (R^2)_{\m\n}+ c_3 (\nabla^2 R)_{\m\n}]+
\cdots = 0 \quad,\ee where $x^\m =(t,x)$. The metric $g_{\m\n}$
should be an effective classical metric associated to the vacuum
state, and a straightforward way to define $g_{\m\n}$ is to take
that the corresponding spatial metric $h_{ij}$ is given by \be
(\det\, h)\, h^{ij}(t,x)= \langle 0 |\hat E_a^i(x) \hat E^{ja}
(x)|0\rangle \quad.\ee However, the main problem with this
approach is to determine the time evolution parameter $t$. It will
be a function of the parameters which appear in the vacuum state,
and it is not obvious how to do this. This is the difficult
problem of the time variable in canonical quantum gravity
\cite{Ish}, and it is related to the problem of the choice of the
scalar product and the normalizability of the $|0\rangle$ state.

A less straightforward, but more promising approach is to use the
De Broglie-Bohm formalism \cite{dbb}. Then the effective classical
metric will be a solution of the quantum equations of motion given
by \be f_i^a(\dot E,E,N_\a ) = {\d S[E]\over \d E^i_a
(x)}\quad,\ee where $f(\dot E,E,N_\a )$ is the expression for the
$\tilde p$ from the canonical formalism, $N_\a$ are the lagrange
multipliers and $S[E]$ is the phase of the wavefunction $\Phi[E]$.
This is a technically simpler approach then the expectation value
approach, and the additional advantages are that there is no
problem with the interpretation of the wavefunction of the
universe\footnote{The wavefunction is interpreted as a wave which
guides the point particles, and hence there is no need for the
external observer because the physical objects and their
properties exist independently of the acts of measurement.} and
there is no need for the specification of a scalar product in
order to verify the semi-classical limit.

In the case when $\L \ne 0$, the Hamiltonian constraint can be
solved in the holomorphic representation via the Kodama
wavefunction \cite{Ko} \be \Psi_K (A) =
\exp\left({1\over\L}\int_\S Tr (A\wedge dA + i\frac23 A\wedge
A\wedge A ) \right)\,. \ee One can then construct the
corresponding state in the spin network basis via the spin network
invariants for the $SU(2)$ CS theory \cite{Smo} \be I_\g = \int
\cd A \, W_\g [A]\,e^{S_{CS}/\L} \quad.\label{cci}\ee When
$\S=S^3$ one can argue that this invariat should be defined as a
Wick rotation of the Kauffman bracket evaluation for the spin
network $\g$. More precisely, the evaluation for $q=e^{2\pi
i/(k+2)}$, where $k\in \bf N$ and $\L l_P^4 =6\pi/k$, is
analytically continued by $k\to ik$ \cite{Smo}.

It has been conjectured that the corresponding state in the spin
network basis corresponds to the de-Sitter vacuum spacetime when
$\L >0$ \cite{Smo}. However, the main problem with this conjecture
is that the Kodama wavefunction is not peaked around any
particular eigenvalue of the $\hat E$ operator, since \be \hat E
\Psi_K = {1\over\L}\hat F \Psi_K \quad.\label{kse}\ee Hence,
exactly as in the $\L=0$ case, $\Psi_K$ cannot be a very good
description the de-Sitter vacuum, where the spatial metric is also
flat.

One can then try to modify the Kodama wavefunction $\Psi_K (A)$ by
multiplying it by a $\Psi_0 (A)$ which is peaked around the $E_0$
eigenvalue, but unlike the $\L=0$ case, the product $\Psi_K
(A)\Psi_0 (A)$ does not solve the Hamiltonian constraint. However,
if one takes \be \Psi_0 (A) = \d \left( F - \L E_0 \right) \quad,
\ee then $\Psi_K \Psi_0$ almost solves the Hamiltonian constraint,
i.e. $\hat C_H \Psi_K \Psi_0 = 0$ when $F\ne\L E_0$, while in the
reduced phase space $(A^*,E^*)$, determined by the constraint $F =
\L E_0 $, one has that $\Psi_K$ is an eigenstate of the $\hat E^*$
operator, see (\ref{kse}). This then suggests to define $I_\g$ via
the expression \be I_\g = \int \cd A \,\d \left( F - \L E_0
\right)
 W_\g [A]\,e^{S_{CS}/\L} \quad,\label{mcci}\ee
where $E_0$ are the flat space triads. It remains to be seen how useful
is this expression for formulating a state sum invariant, but given
such an invariant one can obtain the wavefunction in the triad representation by
using the formulas of the section 5.

\bigskip
\bigskip
\noindent{\bf APPENDIX}

\bigskip
An $SU(2)$ group element $g$ can be represented as \be g
=e^{i\Th^a \s_a}= \cos (\Th/2)\,I + 2i n^a \s_a\, \sin (\Th/2) \ee
where $\s_a$ are the Pauli matrices, $n^a =\Th^a/\Th$ and $\Th^2 =
\Th^a \Th_a$. The trace of the $D^{(j)}(g)$ matrix is given by \be
\chi^{(j)}(g) = {\sin (j + \ha )\Th \over \sin(\Th/2)}\quad. \ee
Since $g_l = e^{iA_l^a \s_a}$, then the gauge invariant plane-wave
coefficients are given by \be C_j (E) = {1\over 32\pi^2}\int_{{\bf
R}^3} d^3 A \left( {\sin (A/2) \over A/2}\right)^2 {\sin (j + \ha
)A \over \sin (A/2)}\, e^{iA^a E_a} \quad,\ee see \cite{fksf}.
This integral can be written as \be {1\over 4\pi}\int_0^\infty dA
\int_0^\pi d\th \sin^2 \th \,\sin (A/2) \sin (j + \ha )A \,
e^{iAE\cos\th}\quad,\ee which becomes an integral \be{1\over 4\pi}
\int_0^\infty dA [\cos(jA) -\cos(j+1)A ][J_0 (AE)+J_2
(AE)]\quad,\ee due to the formula \be J_n (z)
={i^{-n}\over\pi}\int_0^\pi e^{iz\cos\th} \cos(n\th) \quad.\ee

By using \be \int_0^\infty J_n (at) \cos (bt) dt = {\cos(n
\arcsin(b/a))\over\sqrt{a^2 - b^2}}\quad,\quad 0\le b<a \ee and
\be \int_0^\infty J_n (at) \cos (bt) dt = {-a^n \sin(\pi
n/2)\over\sqrt{b^2 - a^2}[b+\sqrt{b^2 - a^2}]^n} \quad,\quad
0<a<b\quad, \ee see \cite{AS}, one obtains (\ref{cj}).

\bigskip
\bigskip
\noindent{\bf ACKNOWLEDGEMENTS}

\bigskip
I would like to thank Nuno Costa Dias and Joao Prata for the
discussions. Work supported by the FCT grants POCTI/FNU/49543/2002
and POCTI/MAT/45306/2002.

\newpage

\begin{figure}[h]
\centerline{\psfig{figure=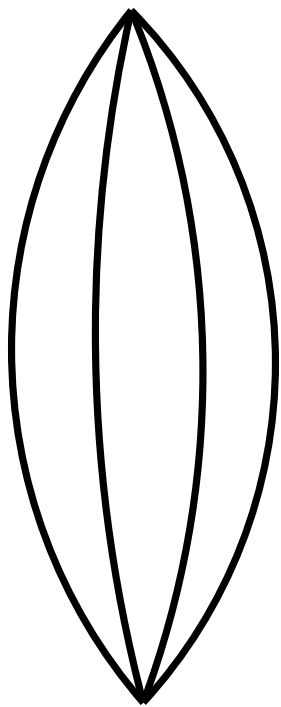}} \caption{The dual one-complex
graph $\G$ of a triangulation of $S^3$ with two tetrahedra.}
\label{one}
\end{figure}

\begin{figure}[h]
\centerline{\psfig{figure=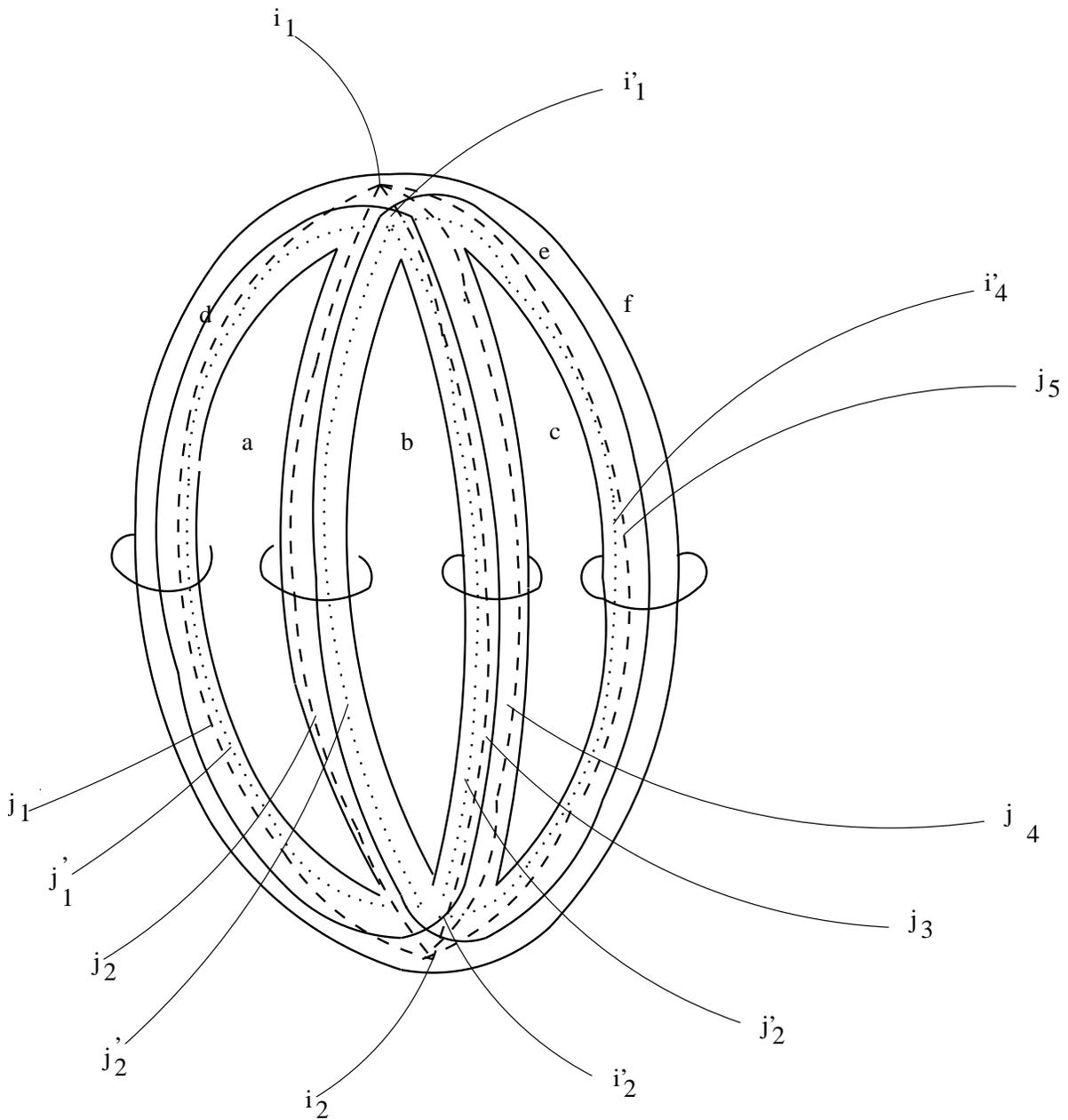}} \caption{The graph
$\G_c(\G,\g)$ for a spin network $\g =\th_5$ embedded in $S^3$
whose one-complex $\G$ is given by that of Fig. 1. The small
circles around the edges of $\G_c$ indicate the group
integrations.} \label{two}
\end{figure}

\begin{figure}[h]
\centerline{\psfig{figure=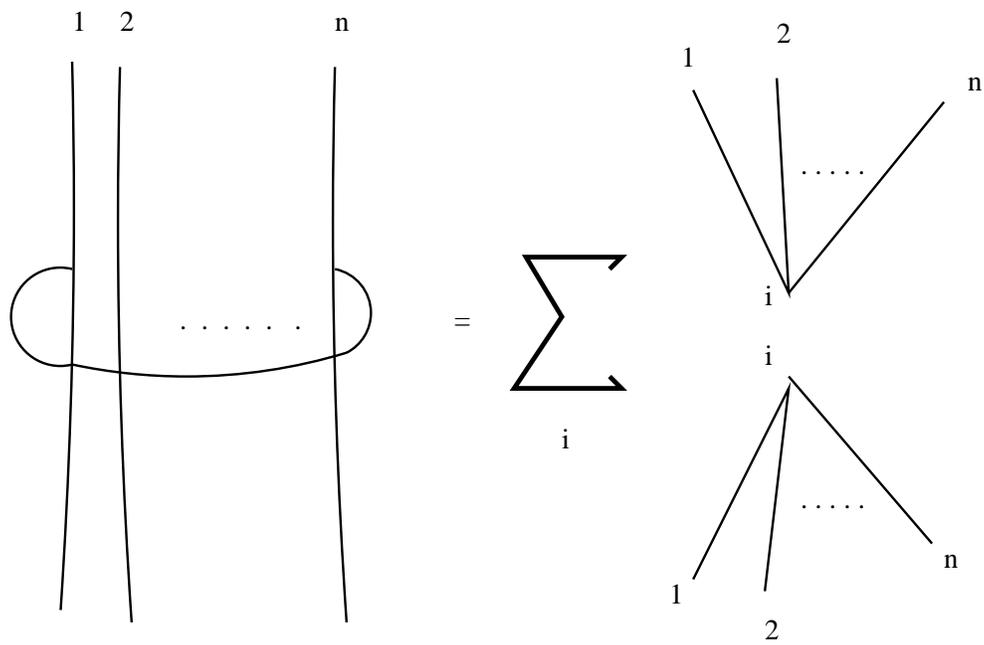}} \caption{The graphical
representation of the group integrations.} \label{three}
\end{figure}

\begin{figure}[h]
\centerline{\psfig{figure=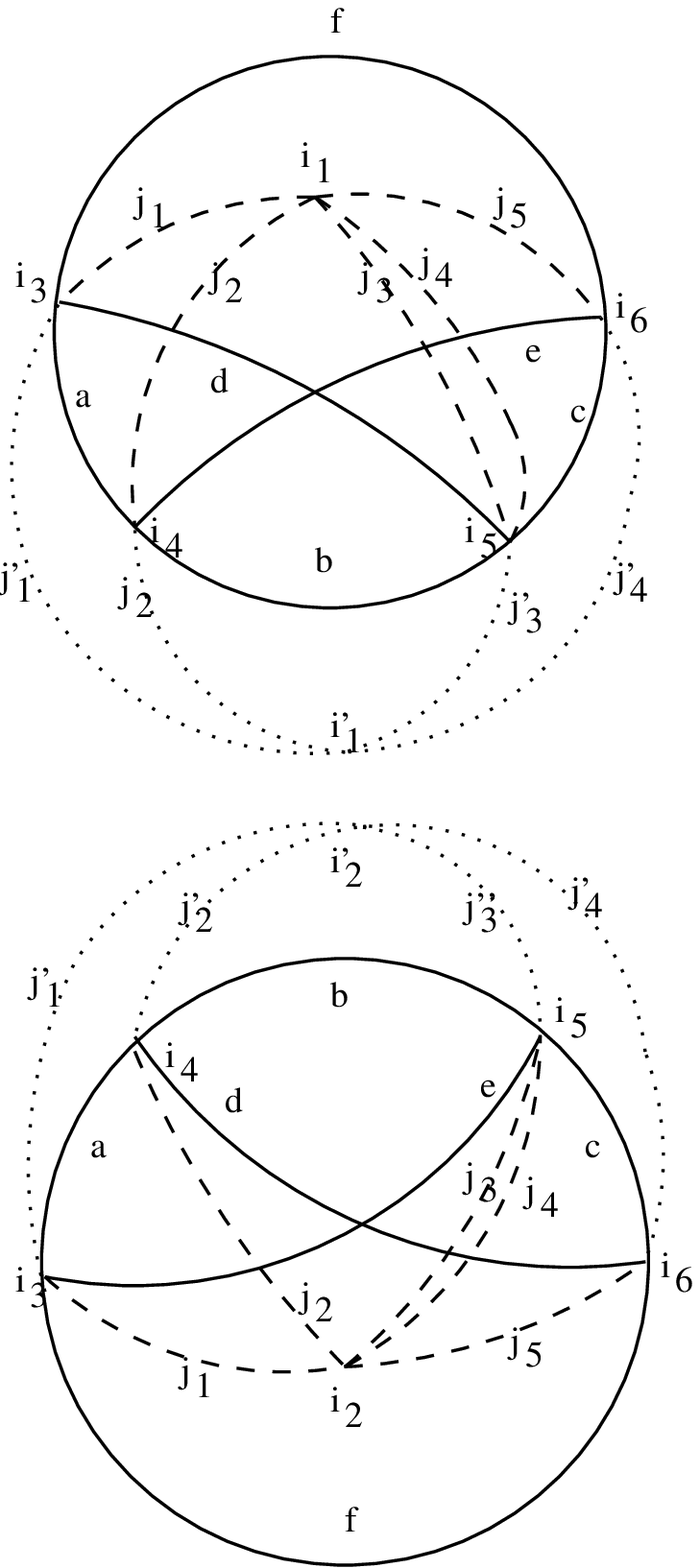}} \caption{The vertex spin
networks for the graph from Fig. 2.} \label{four}
\end{figure}

\begin{figure}[h]
\centerline{\psfig{figure=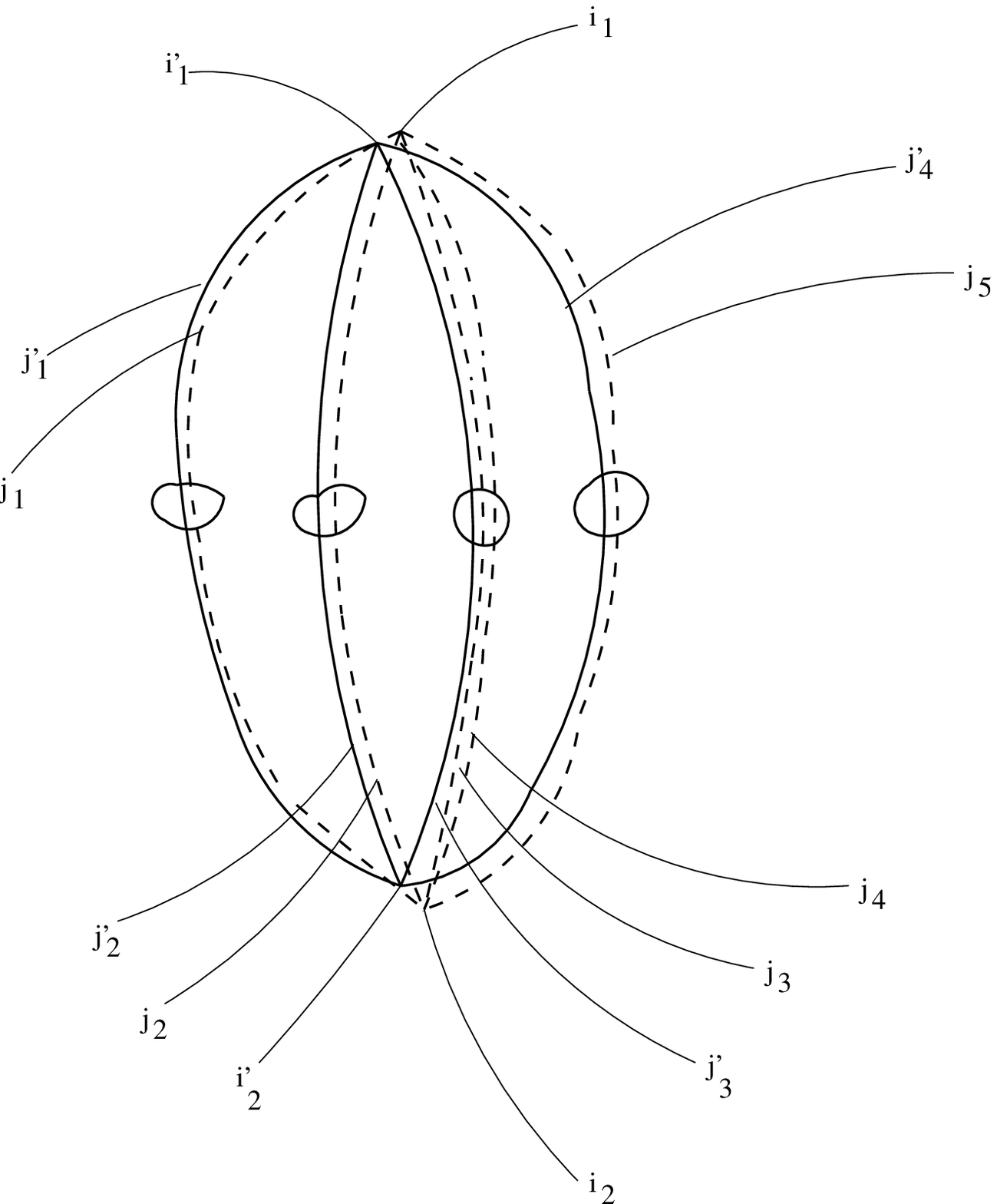}} \caption{The graph
$\G\bigcup\g$ for $\G=\th_4$ and $\g=\th_5$.} \label{five}
\end{figure}

\begin{figure}[h]
\centerline{\psfig{figure=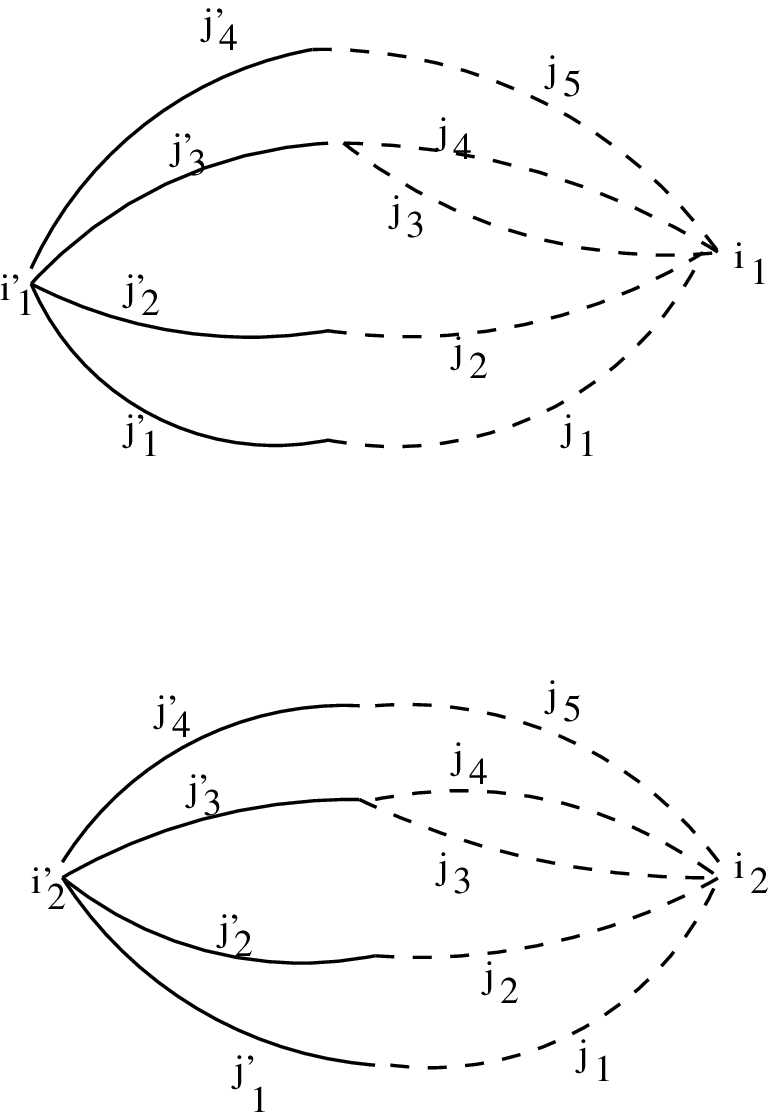}} \caption{The vertex spin
networks for the graph from Fig. 5.} \label{six}
\end{figure}

\end{document}